# Network-Coding Approach for Information-Centric Networking

Muhammad Bilal and Shin-Gak Kang

*Abstract*—**The current internet architecture is inefficient in fulfilling the demands of newly emerging internet applications. To address this issue, several over-the-top (OTT) application-level solutions have been employed, making the overall architecture very complex. Information-centric-networking (ICN) architecture has emerged as a promising alternative solution. The ICN architecture decouples the content from the host at the network level and supports the temporary storage of content in an in-network cache. Fundamentally, the ICN can be considered a multisource, multicast content-delivery solution. Because of the benefits of network coding in multicasting scenarios and proven benefits in distributed storage networks, the network coding is apt for the ICN architecture. In this study, we propose a solvable linear network-coding scheme for the ICN architecture. We also propose a practical implementation of the network-coding scheme for the ICN, particularly for the content-centric network (CCN) architecture, which is termed the coded CCN (CCCN). The performance results show that the network-coding scheme improves the performance of the CCN and significantly reduces the network traffic and average download delay.**

## I. Introduction

RECENTLY, a profound increase in Internet connectivity has been observed. With new emerging Internet applications, such as online social-networking applications, live video streaming, video sharing, multi-user online gaming, and IoT, Internet semantics have changed from host centric to content centric. To meet the requirements of emerging Internet applications, several application-layer solutions are employed in the current Internet architecture, known as over-the-top (OTT) applications such as the content delivery network (CDN), web caching, and peer-to-peer networking [1-6]. However, the additions of new OTT applications are leading towards more complex Internet architecture. Van Jacobson identified a fundamental paradigm shift in Internet services [7] and introduced a fresh concept of Internet architecture, known as information-centric networking (ICN). In the ICN model, "in-network caching" is an integral part of the ICN service framework [8-10]. Unlike CDNs, web caching, and P2P networking, the ICN is a network-layer solution; hence, all ICN-enabled routers are responsible for storing the downloaded content for a limited time. In the ICN paradigm, users request content-by-content names and the network maps the request to the location of the content across the network. The ICN routers create a cache network (CN), wherein cache routers perform caching and other processing operations.

Because of the caching abilities, the network traffic in ICN is significantly less compared to the conventional IP networks. In ICN, the dissemination of content instantly after content publishing can be considered a single source (publisher as a single source) multicast (consumers are multiple receivers) scenario. Though, with the passage of time, the content spreads out in the network in the form of content segments and can be temporarily stored using various intermediate cache routers, termed the custodian nodes. Now, further requests from the consumers can be considered a multisource, multicast scenario. However, If a publisher employs the off-path caching strategy, the content dissemination is considered a multisource, multicast scenario from the start.

Network coding is a widely studied technique for multisource, multicast scenarios [11-15]. In comparison to the conventional multicast solution, which requires group management protocols and construction of multicast trees, the network coding can achieve multicast using a feasible coding scheme computable in polynomial time [11, 12]. Moreover, in conventional IP-based networks, a network-level multisource-multicast solution is unavailable [16]. In contrast, the ICN is a network-level solution, and fundamentally, the ICN can be considered as a multisource, multicast content-delivery solution. Additionally, the network coding has been proven beneficial in other types of content-oriented distributed storage networks [17] such as peer-to-peer [18-20] and CDN [21]. Hence, because of the benefits of network coding in multicasting scenarios and proven benefits in distributed storage networks, the network coding is apt for the ICN architecture.

However, there is very few work has been published related to the practical implementation of network coding in ICN. In [22] authors identified few use cases of network coding in ICN architecture. In [23] authors formulated the network coding problem in ICN as a linear program and presented numerical results to prove the benefits of network coding.

This work is supported in part by research funds 2018, Hankuk University of Foreign Studies, and in part by ICT R&D program of MSIP/ITTP. 2016-0-00192, Standards development for service control and contents delivery for smart signage services.

Muhammad Bilal is with Hankuk University of Foreign Studies, Mohyeon-eup, Cheoin-gu, Yongin-si, Gyeonggi-do, 17035, Rep. of Korea (e-mail: m.bilal@ieee.ac.kr).

Shin-Gak Kang is with Electronics and Telecommunications Research Institute, 305-700 Daejeon, Rep. of Korea (e-mail: sgkang@etri.re.kr).

In [24] authors presented a coded-caching scheme for the ICN, but this short article does not provide details of the proposed scheme. Moreover, in coded-caching, the consumer sends the coefficient vectors for the linear coding, but it is not clear how consumer get this information.

In another study [25], a network coding scheme for an ICN architecture [26], similar to Data-Oriented Network Architecture (DONA) [27] was proposed. The proposed scheme utilized practical fountain codes called Luby transform codes [28], as an information theoretic approach for encoding content, and proved the benefit of network-coding for DONA kind of ICN architecture [26].

In other work [29], the author extended the work of fountain coding scheme for the edge network with an opportunistic network environment. Other work [30], presents a network-coding scheme for the content-centric network (CCN) architecture called NC-CCN. However, NC-CCN scheme violates the basic naming semantics of CCN architecture and makes the pending interest table (PIT) ineffective.

In another study [31], the authors presented a low latency, low loss streaming mechanism, convenient for high-quality delay-sensitive video streaming in CNN/NDN, named L4C2. The L4C2 scheme is based on RLNC model discussed in [32]. In L4C2 requesting nodes first estimate the acceptable delay and probability of packet loss and based on estimation node can target the retrieval of lost coded data packet from cache. Owing to adaptation of RLNC in L4C2, the proposed scheme also provides an efficient way of multicast and multipath delivery of streaming video. However, this scheme is based on assumption that the traffic volume of high-quality delay-sensitive video applications competing with other traffic will increase. This assumption can raise issues of fairness in network.

In [33] authors used the network coding in CCN/NDN architecture to provide an efficient, secure and lightweight authentication scheme. The main objective of proposed scheme is to find the optimal coding assignment that minimizes the total calculation cost while satisfying the required security level.

In this study, first, we show that a linear coding scheme exists for the ICN architecture, which is solvable in polynomial time. We adopted the model and approach from another study [11]. In later sections, we show that the linear-coding techniques can be employed for the ICN and that there are many advantages of using the network coding for the ICN architecture. The remainder of this paper is organized as follows. In section II, we introduce the network coding for the ICN architecture. In Section III, we describe an algebraic approach to the network coding for the ICN. In Section IV, the proposed scheme is described in detail. Section V presents the results of the performance analysis of the proposed scheme against native ICN and IP-based network. Finally, we provide concluding remarks in Section VI..

## II. NETWORK CODING FOR ICN

Caching is an integral part of the ICN architecture. A node that generates content is called a publisher node, and a node that requests content is known as a consumer node; a consumer or publisher node can be a human held device or an automated machine. The published content should be permanently stored in at least one cache node; it can be a publisher node or any other custodian node. The ICN-enabled cache routers can store the content segments for future use; hence, the content is temporarily cached in a few intermediate cache routers while it is being delivered to a consumer. If the content requests traverse a cache router that holds a temporarily cached copy of that particular content segment, the request is entertained locally without being routed to the publisher. Therefore, the delivery of the content to the consumer space in response to further requests can be considered a multisource, multicast scenario.

Fig. 1 shows a simple example of the content delivery in the ICN in the presence of network coding. A publisher P publishes a content object $O_1$, which is divided into two equal-sized segments $seg_1$ and $seg_2$. The intermediate cache routers B and C contain temporary copies of $seg_1$ and $seg_2$, respectively. The cache routers E, F, and I are gateway routers connected to the consumer space. Note that the directed graph shown in Fig. 1 represents the data delivery graph of an actual undirected network. We assume that at some arbitrary time $t_0$, the three gateway routers inject the interest packet for the content $O_1$. In a conventional ICN network, both the segments cannot be delivered to all the three receiver-gateway cache routers because an intermediate link DG is shared between all the paths. However, with a linear-coding scheme, the cache router D helps in linearly combining the segments, and hence, the three receiver-gateway cache routers can receive the data simultaneously.

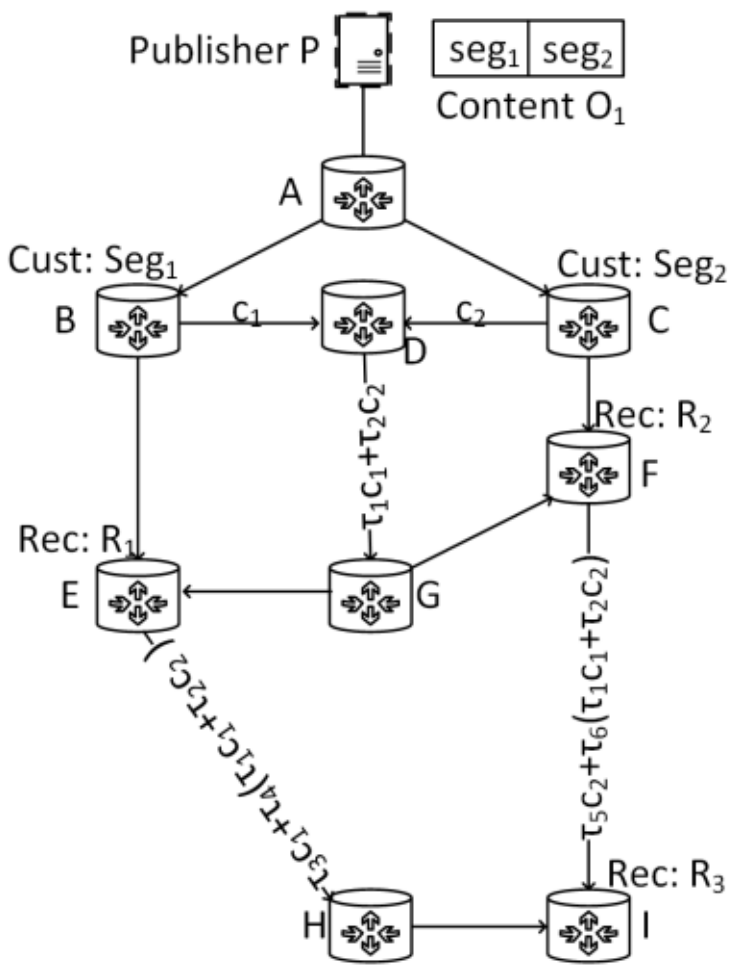

Fig. 1. Simple example of content delivery in the presence of network coding.

The receiver $R_1$ receives $c_1$ (the encoded form of $seg_1$) directly from the custodian cache router B and linearly combined encoded symbol $(c_1, \tau_1 c_1 + \tau_2 c_2)$ from the cache router G. Similarly, $R_2$ and $R_3$ receive $(c_2,\ \tau_1 c_1 + \tau_2 c_2)$ and $(\tau_3 c_1 + \tau_4\ (\tau_1 c_1 + \tau_2 c_2\ ), \tau_5 c_2 + \tau_6(\tau_1 c_1 + \tau_2 c_2))$, respectively. All the encoded symbols are the linear combination of the segments of the requested content. The receiver-gateway routers can retrieve the data by solving a simple system of linear equations (discussed in detail in subsequent sections). The three receivers receive the encoded symbols simultaneously and each receiver is considered to have been allocated all the network resources. However, three important questions need to be answered. 1) Does the linear-coding scheme exist for the ICN architecture? 2) Under what conditions does the polynomial-time solvable linear-coding scheme exist for the ICN architecture? 3) How to implement a linear-coding scheme for the ICN architecture? In the subsequent section, we answer these questions.

### A. Coding Theorem for ICN

We consider a scale-free acyclic graph G = (V, E), where V is a set of vertices representing the cache router, and E is a set of edges. The edge capacity is assumed to be one content segment per unit time, i.e., 1 seg/s, where each segment is considered to be $m$ bits long. The high-capacity links between the cache routers are defined as multiple parallel unit capacity edges in G. Further, a cache router stores one or multiple segments of the same content object based on a cache-replacement policy.

**Definition:** Min-cut is the set of minimal number of edges between two vertices in G, the removal of which is sufficient to disconnect the vertices. Min-cut also defines the number of disjoint paths between two vertices.

Note that in our model, a link with higher capacity is defined as multiple parallel unit capacity edges in G; therefore, in a min-cut, the link with higher capacity is counted multiple times.

The min-cut max-flow (MCMF) is a well-studied theorem in the field of network coding [11, 12]. The MCMF theorem states that for a graph G = (V, E) with unit capacity edges, the min-cut between any two vertices is exactly equal to the number of disjoint paths. As discussed earlier, the content delivery in the ICN can be considered a multisource, multicast scenario. Accordingly, we redefine the MCMF theorem for the ICN as follows.

**Theorem 1:**

Consider a graph G = (V, E) with unit capacity edges. At any given time, we can transform $G \rightarrow G' = (V' \equiv V \cup V_s, E' \equiv E \cup E_s)$, where $G'$ represents the extension of graph $G$ with an identified multiple independent or linearly correlated publisher and custodian nodes represented by $V_s \subset V'$. The set of receiver-gateway cache routers $V_R$ can receive data from all source nodes $V_s$ simultaneously with at most $m(min(V_S \leftrightarrow V_R)\ mincut)$ bits/s. This simultaneous transmission exists over a sufficiently large finite field $F_{2^m}$, wherein some or all intermediate nodes are used to linearly combine the information, and the coefficients of the linear coding are chosen independently and uniformly over the finite field $F_{2^m}$.

**Proof:**

In the ICN, the information received by the gateway-cache router $R_j$ is originated from multiple source nodes $V_s$. With the unit capacity edges, the max-flow is equal to $V_s \leftrightarrow R_j$ min-cut or number of disjoint paths. Hence, using the MCMF theorem, the possible maximum traffic flow between $R_j$ and source nodes $V_s$ is equal to the weighted sum of the $V_s \leftrightarrow R_j$ min-cut. Hence, the simultaneous transmission $\forall R_j \in V_R$ is the weighted sum of the minimum $V_s \leftrightarrow V_R$ min-cut. This is possible because an overlapping edge of all disjoint paths linearly combines the data packets. Moreover, based on [34, theorem 1], there always exists a positive probability of the success of the simultaneous transmission, which is directly proportional to the size of the finite field $F_{2^m}$ and inversely proportional to the path length.

## III. Algebraic Approach to Network Coding for ICN

Network coding has various theoretical frameworks [11-12, 29-31]. In this study, we employed an algebraic approach to network coding, as discussed in another study [11].

We reconsider the example shown in Fig. 1. In the given example, the cache routers D, E, and F are the best points for performing the linear coding on the incoming data packet. Further, we assume that the set of values of coefficients $\{\tau\}$ of the linear coding is defined over a sufficiently large finite field $F_{2^m}$ such that coding can be performed (Theorem 2). The coding points and cache routers D, E, and F associate a coefficient value to all incoming edges. The set of coefficients associated with each coding point is called the coding vector. For a given example, we can define the transform matrices $T_1, T_2$, and $T_3$ for $R_1, R_2$, and $R_3$, respectively, as given below.

$$T_1 = \begin{bmatrix} 1 & 0 \\ \tau_1 & \tau_2 \end{bmatrix},\ T_2 = \begin{bmatrix} 0 & 1 \\ \tau_1 & \tau_2 \end{bmatrix}, \text{ and } T_3 = \begin{bmatrix} \tau_{3+}\tau_1\tau_4 & \tau_2\tau_4 \\ \tau_1\tau_6 & \tau_5 + \tau_6\tau_2 \end{bmatrix}$$

We consider $W_{l,k}$ as the symbol received by the gateway-cache router $R_j$ at the $l$th incoming edge for content $O_i$. The receiver can then recover all $K$ encoded symbols (simultaneously sent by $K$ sources) by solving the following linear system of equation.

$$\begin{bmatrix} c_1 \\ c_2 \\ .. \\ c_K \end{bmatrix} = T_j^{-1} \begin{bmatrix} W_{l,1} \\ W_{l,2} \\ .. \\ W_{l,k} \end{bmatrix}$$

In a coded network, for a receiver $R_j$, the network is an $in(R_j)$ output system, where $in(R_j)$ represents the number of incoming edges. Moreover, the receiver $R_j$ can decode the incoming encoded symbols using a network transformation matrix $T_i$. To route the requested content segments simultaneously from multiple sources $V_s$ to a receiver-gateway cache router $R_j$, we have to obtain $size(V_s)$ number of disjoints that connect all the sources. Hence, for multiple receiver-gateway cache routers $V_R$, we have $size(V_R)$ number of disjoint paths. These paths may overlap on certain edges. To ensure a simultaneous reception, the tail nodes of the overlapping edges are the best coding points in the network. An equivalent algebraic form of Theorem 1 is given below.

**Theorem 2:**

Consider a graph G = (V, E) with unit capacity edges. At any given time, we can transform $G \rightarrow G' = (V' \equiv V \cup V_s, E' \equiv E \cup E_s)$, where $G'$ represents the graph G with an identified multiple independent or linearly correlated publisher and custodian nodes represented by $V_s \subset V$. The set of receiver-gateway cache routers $V_R$ can receive data from all source nodes $V_s$ simultaneously with at most $m(min(V_S \leftrightarrow V_R)\ mincut)$ bits/s. This simultaneous ~~transmission~~ exists if the coding vectors $\{\tau_k\}$ are selected over a sufficiently large finite field $F_{2^m}$, such that all the matrices of the gateway routers $T_j$ are full rank.

**Proof:**

Based on our assumption that G = (V, E) is acyclic scale- free and that the intermediate nodes are allowed to perform only linear coding, each entry in $T_j$ is polynomial in $\{\tau_k\}$, where $\{\tau_k\}$ is defined over a large finite field $F_{2^m}$. The proof strategy of Theorem 2 is to show that the product of the determinant of the transform matrices of the gateway routers is non zero, i.e., $\prod_j \det(T_j)\ \forall\, j = 1,2,3..N$, where $N$ is the total number of receiver-gateway cache routers. As each entry in the $T_j\ \forall j = 1,2,3..N$ is polynomial in $\{\tau_k\}$, we can infer that $\prod_j \det(T_j)$ is also a polynomial.

According to the Schwartz–Zippel Lemma, we have the following:

$$P\left[\prod_j \det(T_j) = 0\right] \leq \frac{d}{|2^m|}$$

where $d$ is the degree of polynomial $\prod_j \det(T_j)$, and $P\left[\prod_j \det(T_j) = 0\right]$ is the probability that polynomial $\prod_j \det(T_j)$=0. Hence, we conclude that over a sufficiently large finite field $F_{2^m}$, a set of coefficients $\{\tau_k\}$ exist that ensures that all the matrices $T_j$ are full rank. Further, the probability of obtaining feasible coefficients becomes higher for larger values of $m$.

## *A. Algebraic Framework*

From Theorems 1 and 2, we can conclude that for the ICN type of cache network, a linear coding scheme exists. Next, we will obtain the network transformation matrix $T_j$ for each gateway cache router.

Fig. 2 shows the equivalent line graph of example shown in Fig. 1. An edge between any two vertices in a line graph represents a cache node and an outgoing edge in the corresponding network graph G = (V, E). For instance, in Fig. 2, an edge $DG \rightarrow GF$ represents the cache router $G$ and outgoing edge from $G \rightarrow F$ in the network-graph representation of Fig. 1. Hence, we can conclude that an edge in a line-graph representation of the ICN cache network is a memory with a size equal to the size of the cache router, which is common for two adjacent vertices in the line graph. Moreover, each vertex in the line graph represents the transitional memory, which stores the transmitted encoded symbols on the edge for $d^p_{v,u}$ time duration, where $d^p_{v,u}$ is the propagation delay between two adjacent vertices in the network graph G = (V, E). If there are $size(V_s)$ number of sources (publisher and custodian cache routers) each transmitting 1 seg/s and if intermediate routers are allowed to perform linear coding, the line graph can be considered a $size(V_s)$ input and $size(V_s)$ output linear system, which can be represented by the system of linear equations.

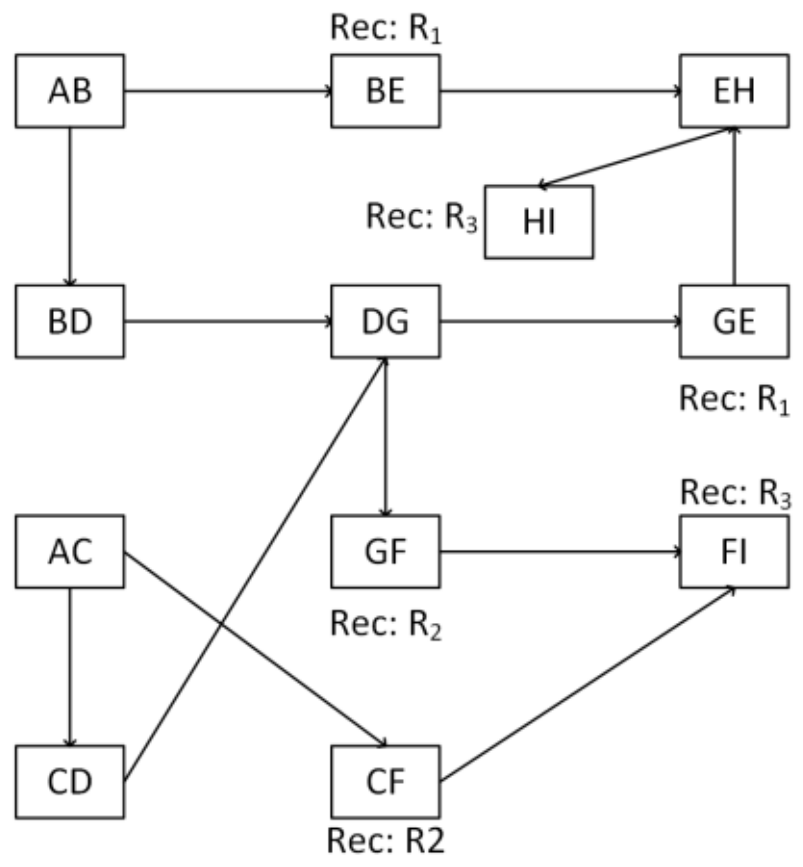

Fig. 2. Line graph representation of example 1.

Further, according to Theorem 2, the system of linear equations is solvable at the receiver-gateway router (and at any intermediate cache router) if the coefficients $\{\tau_k\}$ of the linear equations are chosen from a sufficiently large finite field $F_{2^m}$. Using an approach similar to that given in another study [11], we derive the system of linear equation for this system as follows. At the coding point, the cache router performs the following linear operation.

$$c_j = \sum_{\{i:seg_i \in v \text{ or } v \text{ is a publisher}\}} b_{i,j} seg_{j,i} + \sum_{\{k \in in(v)\}} \tau_{k,j} c_k$$

Following the above process, the final output process observed by the receiver-gateway cache router $R_l$is given as follows.

$$W_l = \sum_{\{k \in in(l)\}} h_{l,j,i,k} c_k$$

The coefficients $\{b_{i,j}, \tau_{k,j}, h_{l,j,i,k}\}$ are considered in the finite field $F_{2^m}$. To understand the above process, we consider a simple example, as shown in Fig. 3.

After receiving the encoded symbols $c_1$and $c_2$ of same $O_j$, the node $v$ performs the coding operation. In Fig. 3a, the cache router is custodian of $seg_{j,0}$ of content object $O_j$; hence, it linearly combines the incoming encoded symbols along with $seg_{j,0}$. The $seg_{j,0}$ can be considered a random process generated on node $v$, which is combined with the incoming random processes $c_1$and $c_2$. In Fig. 3b, the cache router is not a custodian of any segment of the content object $O_j$. Hence, it linearly combines only the incoming encoded symbols. As discussed earlier, with $size(V_s)$ number of sources, the network $G = (V, E)$ can be considered $size(V_s)$ input $size(V_s)$ output system linear equations. Hence, the aforementioned equations at the network level can be written in the form of matrices.

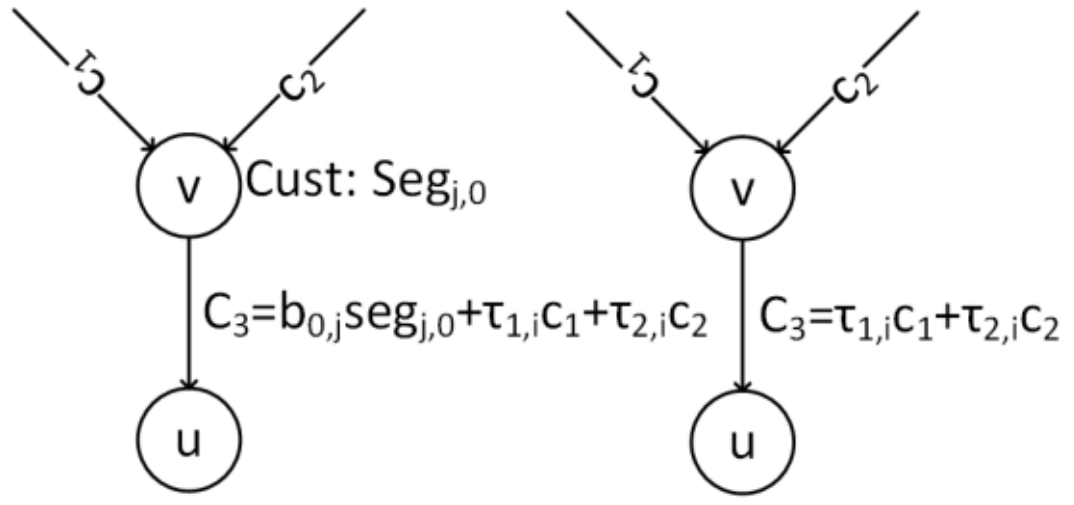

Fig. 3. a) Coding operation when intermediate cache router is a custodian node, and b) Coding operation when intermediate cache router is not a custodian node.

$$C_{k+1} = B\, Seg_k + T\, C_k$$
$$W_l = H_{l,j,i,k} C_k$$

where $C_k$ is a $E \times 1$ matrix representing the encoded symbols for all edges in the network. $B$ is a $size(V_s) \times E$ matrix, which is common for all $R_j$ requested for the same content object. The matrix B represents how the publisher and custodian nodes are connected to the network. $T$ is an $E \times E$ adjacency matrix of the line graph. If an edge in the line graph has a common vertex from the network graph, and the common vertex is a receiver for the first node and a source for the second node, the entry in $T$ is an arbitrary positive value $\tau_{k,j}$, otherwise, it is 0. The matrix $H$ is a $size(V_R) \times E$ matrix and represents how the gateway router $R_j \in V_R$ observes the output process received by the incoming links. Finally, $Seg_k$ and $W_l$ are $size(V_s) \times E$ matrices representing the input and output of the entire system, respectively.

### *B. Transformation Matrix of the Network*

We consider network $G = (V, E)$ as a black box wherein an input process $seg_{j,i}$ is defined by the injection of content segments from the publisher and custodian nodes, and out process $W_l$ is defined by the sequence of encoded symbols observed by the receiver gateway router $R_l$. The transformation of $seg_{j,i} \rightarrow W_l$ requires $L$ number of stages, where $L$ represents the path length between the source and receiver cache routers. In other words, the transform matrix is considered at each stage in the path; thus, at the $Lth$ stage we obtain $I + T + T^2 \ldots T^L$. Because the matrix $T$ is nilpotent, $(I - T^{-1}) = I + T + T^2 \ldots T^L$. This further confirms that the entries in $T_j$ are polynomial in variable $\{\tau_k\}$ chosen from sufficiently large finite field $F_{2^m}$. Thereafter, using the standard linear system theory, the transformation matrix for the gateway router $R_j \in V_R$ is given as follows.

$$T_j(D) = H(D^{-1}I - T)B$$

where matrix $D$ is a $E \times E$ matrix representing the delay observed on each edge in the network $G = (V, E)$. If we consider unit delay, i.e., $d_{v,u}^p = 1$, the following equation is true.

$$T_j = H(I - T)B$$

## IV. Practical Implementation of Network Coding in CCN

In previous sections, from Theorems 1 and 2, we concluded that for the ICN type of cache network, a linear coding scheme exists. In addition, we obtained the network transformation matrix $T_j$ for each gateway-cache router. This section presents our scheme for the practical implementation of the network coding in the ICN, particularly for the CCN architecture, termed the coded CCN (CCCN).

In the conventional CCN architecture, an interest packet is referred to an entire content object or segment of content object, and the payload in the data packet is a segment of the requested objects or entire object itself. However, because of the coding process, the payload of a data packet may have an encoded symbol generated with two or more segments of the requested content object. To implement the network-coding scheme, we introduce few new fields in the interest and data packets.

Fig. 4 shows a simple interest and data packet format of the CCN. All the fields are self-explanatory; however, refer to another study for further details [38]. We introduced new fields, as listed in Table I, in the interest and data packets at "optional

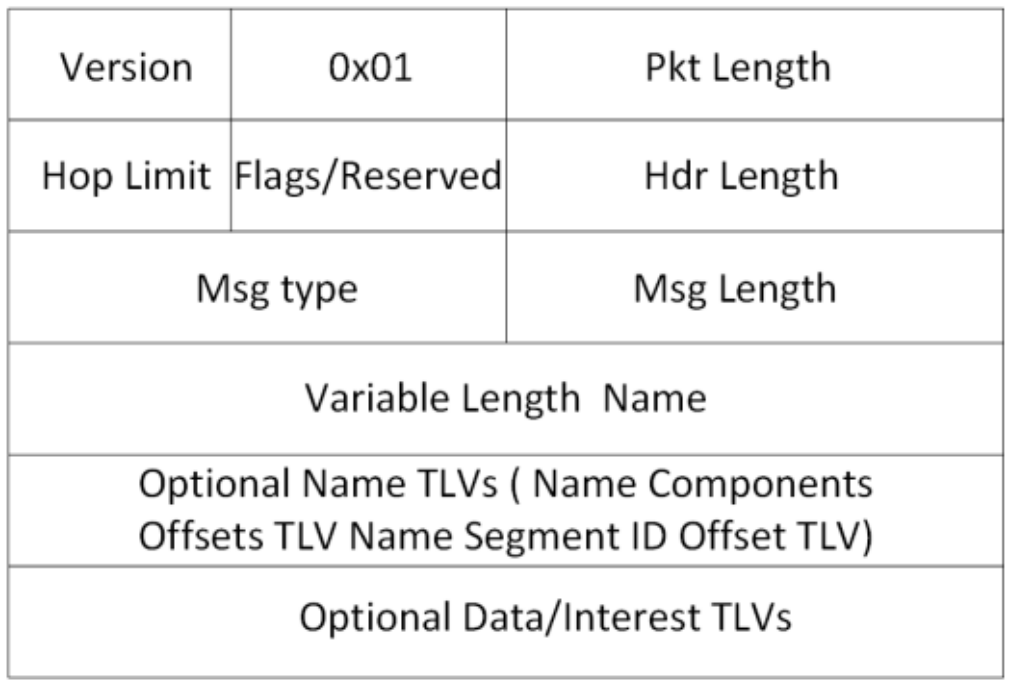

Fig. 4. Interest and data packet format for CCN

interest type-length-values (TLVs)" and "optional data TLVs" fields, respectively.

TABLE I.

| Interest Packet | | | Data Packet | | |
|---|---|---|---|---|---|
| Field Name | Value | Description | Field Name | Value | Description |
| $I_{Type}$ | 0 | The interest packet is intended for the entire content object. | $D_{Type}$ | 0 | The payload is simple data. |
| | −1 | The interest packet is intended for the set of segments of the requested content object. | | −1 | The payload is encoded data. |
| | x | The interest packet is intended for the segment x of the requested content object. | $D_{info}$ | - | It is a variable length string and represents the IDs of the set of segments of the encoded data. This field is used if $D_{Type} = -1$. |
| $I_{info}$ | - | It is a variable length string and represents the IDs of the set of segments of the requested content object. This field is used if $I_{Type} = -1$. | | | |

- $in(v)$ : It returns the set of incoming links on cache router $v$
- $getinf(val)$:
  - For $val = I_{Type}$, value of $I_{Type}$ is returned and if $I_{Type} = -1, I_{info}$ string is returned.
  - For $al = D_{Type}$, value of $D_{Type}$ is returned

and for $D_{Type} = -1$ , $D_{info}$ string is returned.

- For $val = I_{name}$, the name of the requested content object is returned.
- For $al = D^i_{name}$, the name of the content object of the arrived data packet on edge $e_i \in in(v)$ is returned.
- For $val = size$, the size of the requested content object in terms of number of segments is returned.
- For $val = seg^i_{id}$, the segment number of the arrived data packet on edge $e_i \in in(v)$ is returned

- $|v|$: The set of content segments stored in cache router $v$.
- $A_{req}$: The arrived interest packet
- $modify(A_{req})$: It modifies $I_{info}$ string of interest packet $A_{req}$.
- $A^i_{pkt}$: The arrived data packet on $e_i$
- $\{A_{pkt}\}$: A set of data packets belongs to the same content object. Following the similar notation, $|A_{pkt}|$ represents the size of set $\{A_{pkt}\}$.

**Algorithm 1: (Interest handling)**

if $getinf(I_{Type}) = 0$
    if $getinf(I_{name}) \in |v|$
        reply with the requested content
        if $getinf(size) \rightarrow Mutiple\ Segments$
            $modify(A_{req})$ // modifiy $I_{info}$ string
            Forward based on FIB
        end
    else if $getinf(I_{name}) \in |v| \wedge$ encoded //the encoded form of segment is stored in $v$
        reply with encoded segments
        Forward based on FIB
    else
        Forward based on FIB
    end
else if $getinf(I_{Type}) = -1$
    if $getinf(I_{name}) \in |v|$
        reply with the available segments
        if all $I_{info} \notin |v|$ //All the requested segments are not available in local cache
            $modify(A_{req})$ // modifiy $I_{info}$ string
            Forward based on FIB
        end
    else if $getinf(I_{name}) \in |v| \wedge$ encoded //the encoded form of segment is stored in $v$
        reply with encoded segments
        Forward based on FIB
    else
        Forward based on FIB
    end
else if $getinf(I_{Type}) = x$
    if $getinf(I_{name}) \in |v|$
        reply with the available segment
        Forward based on FIB
    else if $getinf(I_{name}) \in |v| \wedge$ encoded //the encoded form of segment is stored in $v$
        reply with encoded segment
        Forward based on FIB
    else
        Forward based on FIB
    end
end

**Algorithm 2: (Coding decision)**

if $getinf(D^i_{name})$ $and$ $getinf(seg^i_{id})$ $are$ $same$ $\forall$ $\{A_{pkt}\}$ $\wedge$ $getinf(D^i_{name}) \in PIT$
    if $\{A_{pkt}\}$ arrive on different $e_i \in in(v)$
        $A^j_{pkt} = \sum_{\{k \in in(v)\}} \tau_{k,j} A^k_{pkt}$
        $Replacement(LFRU)$
        Forward based on routing table
    else if $\{A_{pkt}\}$ arrive on same $e_i \in in(v)$
        $Replacement(LFRU)$
        Forward based on routing table
    else
        Forward based on routing table
    end
else if $getinf(D^i_{name})$ $and$ $getinf(seg^i_{id})$ $are$ $same$ $\forall$ $\{A_{pkt}\}$ $\wedge$ $getinf(D^i_{name}) \in |v|$
    if $\{A_{pkt}\}$ arrive on different $e_i \in in(v)$
        $A^j_{pkt} = \sum_{\{i: seg_i \in |v|\}} b_{i,j} seg_{j,i} + \sum_{\{k \in in(v)\}} \tau_{k,j} A^k_{pkt}$
        $Replacement(LFRU)$
        Forward based on routing table
    else if $\{A_{pkt}\}$ arrive on same $e_i \in in(v)$
        $Replacement(LFRU)$
        Forward based on routing table
    else
        Forward based on routing table
    end
else
    Forward based on routing table
end

**Algorithm 3: (Processing on receiver gateway cache router)**

if $getinf(D_{Type}) = 0$
    $Replacement(LFRU)$
    Forward data in consumer space
else if $getinf(D_{Type}) = -1$
    $W_l = \sum_{\{k \in in(l)\}} h_{l,j,i,k} c_k$// Use the transformation matrix
    $Replacement(LFRU)$
    Forward data in consumer space
    end
end

**Algorithm 1** explains how a cache router handles the interest packet in the CCN when the network nodes are allowed to perform linear coding. A brief stepwise explanation is given below.

CASE 1: If the interest packet is the request for the entire content object, the condition whether the requested content or segment of the content is already stored in the cache is checked. If the content segment(s) is available in the cache, the cache router replies with the available segment(s). If there are missing segments, which are unavailable in the local cache, the cache router modifies the interest packet to search for the rest of the segments of the content.

CASE 2: Similarly, if the interest packet is the request for a particular segment or a set of segments of the content object, the condition whether the segment(s) of the content is already stored in the cache is checked. If the content segment(s) is available in the cache, the cache router replies with the available segment(s). If there are missing segments, which are unavailable in the local cache, the cache router modifies the interest packet to search for the rest of the segments of the content.

CASE 3: However, in both cases (cases 1 and 2 discussed above), if the stored segment(s) is encoded and there are missing segment(s), which are unavailable in the local cache, the cache router replies with the encoded segment(s) and forwards the interest packet without modification to search for the rest of the segments of the content.

**Algorithm 2** explains how a cache router handles the incoming packets when the incoming packet requires encoding and when the packets should be forwarded without the linear coding. A brief stepwise explanation is given below.

CASE 1: If multiple data packets arrive at different edges of the cache router and the request for the arriving contents is in PIT, the cache router then performs the coding procedure $A^{j}_{pkt} = \sum_{\{k \in in(v)\}} \tau_{k,j} A^{k}_{pkt}$ and applies the LFRU [39] cache-replacement scheme to decide whether the encoded segments should be stored. Finally, the encoded segments are forwarded based on the routing table.

CASE 2: Similarly, if multiple data packets arrive at different edges of the cache router and the segment or multiple segments of the same content object were already stored in the cache, the cache router performs the coding procedure $A^{j}_{pkt} = \sum_{\{i:seg_i \in |v|\}} b_{i,j} seg_{j,i} + \sum_{\{k \in in(v)\}} \tau_{k,j} A^{k}_{pkt}$ and applies the LFRU [39] cache-replacement scheme to decide whether the encoded segments should be stored. Finally, forward the encoded segments based on the routing table.

CASE 3: Similarly, if multiple content segments arrive at the same edge of the cache router, the LFRU [39] cache-replacement scheme is used to decide whether the encoded segments should be stored and the encoded segments are forwarded based on the routing table.

**Algorithm 3** explains how the gateway-cache router processes the incoming packets. A brief stepwise explanation is given below.

CASE 1: If the incoming data packet payload is a simple content segment, the LFRU [39] cache-replacement scheme is used to decide whether the segment should be stored, and the segment is then transferred into the consumer space.

CASE 2: If the incoming data packet payload is an encoded content segment, the transformation matrix is used to decode the encoded symbol and the LFRU [39] cache-replacement scheme is used to decide whether the segment should be stored, and the segment is then transferred into the consumer space.

## V. Performance Analysis

### A. *Network Setup and Assumptions*

We consider a scale-free network of 100 cache nodes generated using the Barabási–Albert (BA) model, as shown in Fig. 5, which connects the publisher and the consumer space. Each cache router has a static request routing table. Further, we assume five content publishers in the network, each with 10000 content items; a Zipf-distribution with a popularity distribution exponent $\propto = 0.7$ is used to determine the population of 50000 content items in the entire network.

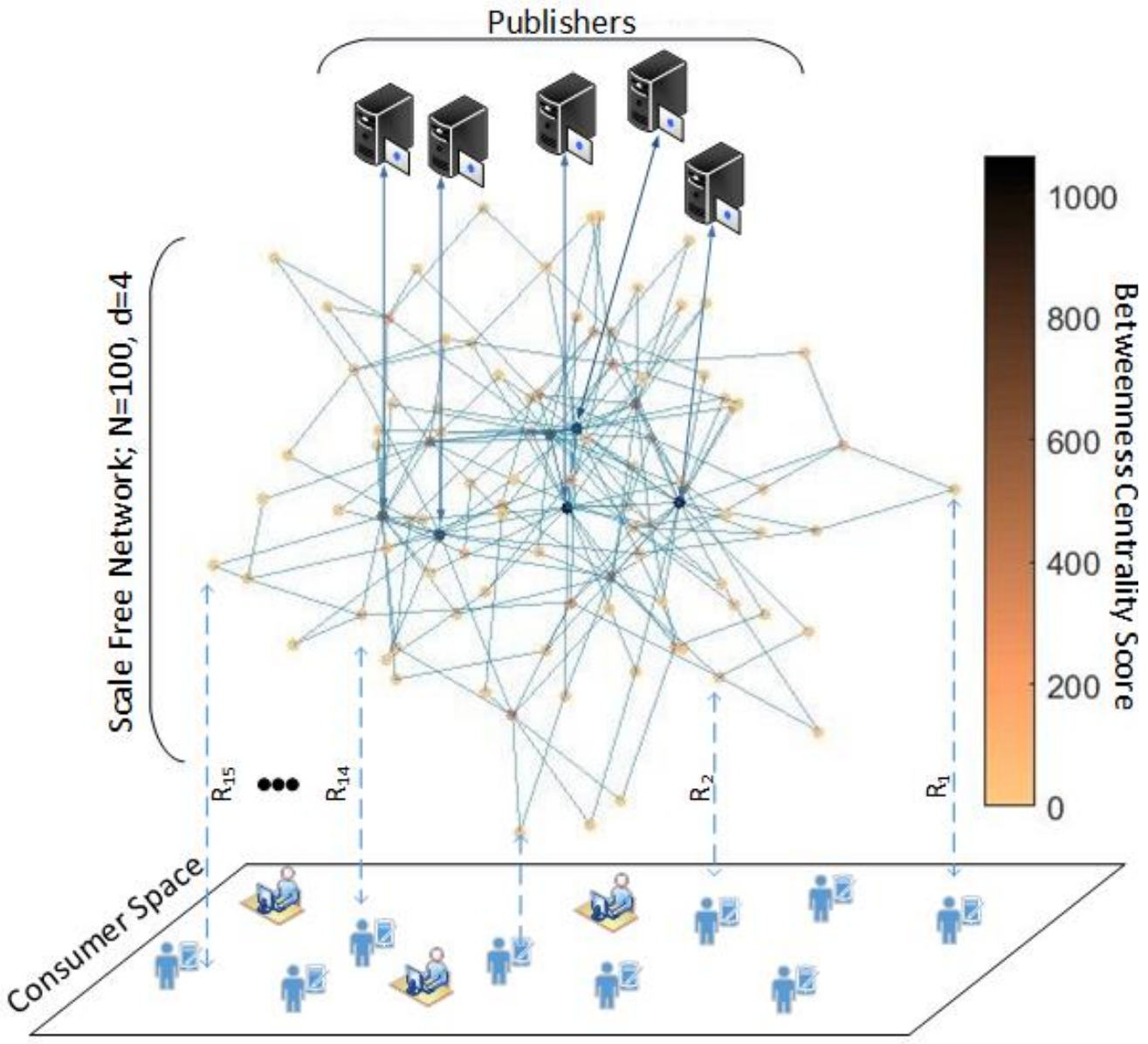

Fig. 5. A depiction of network setup used for the analysis and simulation.

To ensure quick dissemination of the content in the network, the publishers are connected to the cache routers with the highest betweenness centrality score. In Fig. 5, the darker the color of the dot, the higher the betweenness centrality score of the cache node. Furthermore, we assume 15 gateway-cache routers connected to a large number of consumers. In Fig. 5, the dotted line represents the aggregated content request arrival at a gateway-cache router directly from the consumer space. The total request arrival rate ($\lambda_j$) at any gateway cache router $R_j$ is obtained as the sum of the aggregated content requests directly coming from the consumer space ($\lambda^d_j = 25{\sim}100/s$) plus aggregate requests forwarded by the neighboring cache nodes ($\lambda^f_j = \sum \lambda_{i,j}$, where $i \in$ set of adjacent cache nodes). Moreover, the gateway-cache routers $R_j$, which connect the consumer space with the core network, are selected with the lowest centrality score; this ensures that the gateway-cache router can use most of the processing resources for the decoding operation. This assumption is reasonable because on an average, the edge routers in a scale-free network have lower centrality score.

Further, we assume that the size of each content item is 100 MB, which is divided into 10 segments each with a size of 10 MB. The link capacity in the core network between the two cache routers is 1 Gbps, which implies that each link represents 100 parallel unit capacity edges in an equivalent network graph (as discussed in previous sections) of the core network. Finally, the LFRU [39] is used in the experiment as a cache-replacement scheme. We assume that for the LFRU [39] scheme, 20% of the cache is allocated for the unprivileged partition.

### *B. Results*

We implemented the network setup, as described above, in MATLAB and compared the performance of the proposed CCCN scheme with the CCN, NC-CCN [30], and IP-based network. This comparison is made in terms of three performance parameters: the average link-capacity usage, average download time, and average interest load. To incorporate the coding cost in the CCCN and NC-CCN cache routers, we assume that on an average, a cache router takes 5% more time to process a data packet. However, the exact cost of the coding operation is still debatable. Moreover, we consider 0.5-1% packet lost rate, which is consider an acceptable lost rate for high quality video streaming.

Fig. 6 shows the comparison of the average download time observed when each of the 15 gateway-cache routers receives the requests for the contents assumed to be Poisson distributed with a rate $\lambda_j^d = 100\, req/s$.

**Definition:** In the steady state, the average download time is defined as the ratio of the total number of requests observed on all 15 cache routers to the time taken to receive all the requested contents at the gateway routers.

The results are considered for different cache sizes [200 MB–100 GB]. Fig. 6 clearly shows that the CCN, NC-CCN, and CCCN perform better than the IP based network. For a smaller cache size, the CCN performs slightly better than NC-CCN and CCCN; however, for larger cache sizes, the CCCN performs much better than NC-CCN and CCN. The difference in the performances reduces with very large size caches. In fact, with a smaller cache size, several popular contents do not get the opportunity to be stored in the intermediate cache router. This implies that in NC-CCN and CCCN, the intermediate cache routers performs several coding operations, and hence, induces extra delay. For a large cache size, the advantages of coding are much more than the coding cost, and both NC-CCN and CCCN performs better than the CCN. However, for very large cache sizes, the difference in performances reduces because a large number of popular contents are available at the first few hop distance from the gateway router.

The NC-CCN network-coding scheme does not support the segment level naming; rather the content level naming is used for requesting the segment level coded blocks. As a result, the PIT become ineffective, and segment level data search is not possible; therefore, to distinguish different interest and data packets NC-CCN introduced a tagging scheme. The tagging scheme put an extra computational and storage burden on the intermediate cache routers, which introduces additional processing delay. Moreover, NC-CCN defines a priority scheme for the content replacement based on the number of coded blocks stored in the cache router. This kind of content replacement does not ensure that the replicated content is popular in the locality of cache node. All these factors degrade the performance of NC-CCN, as shown in Fig. 6, CCCN outperforms the NC-CCN scheme.

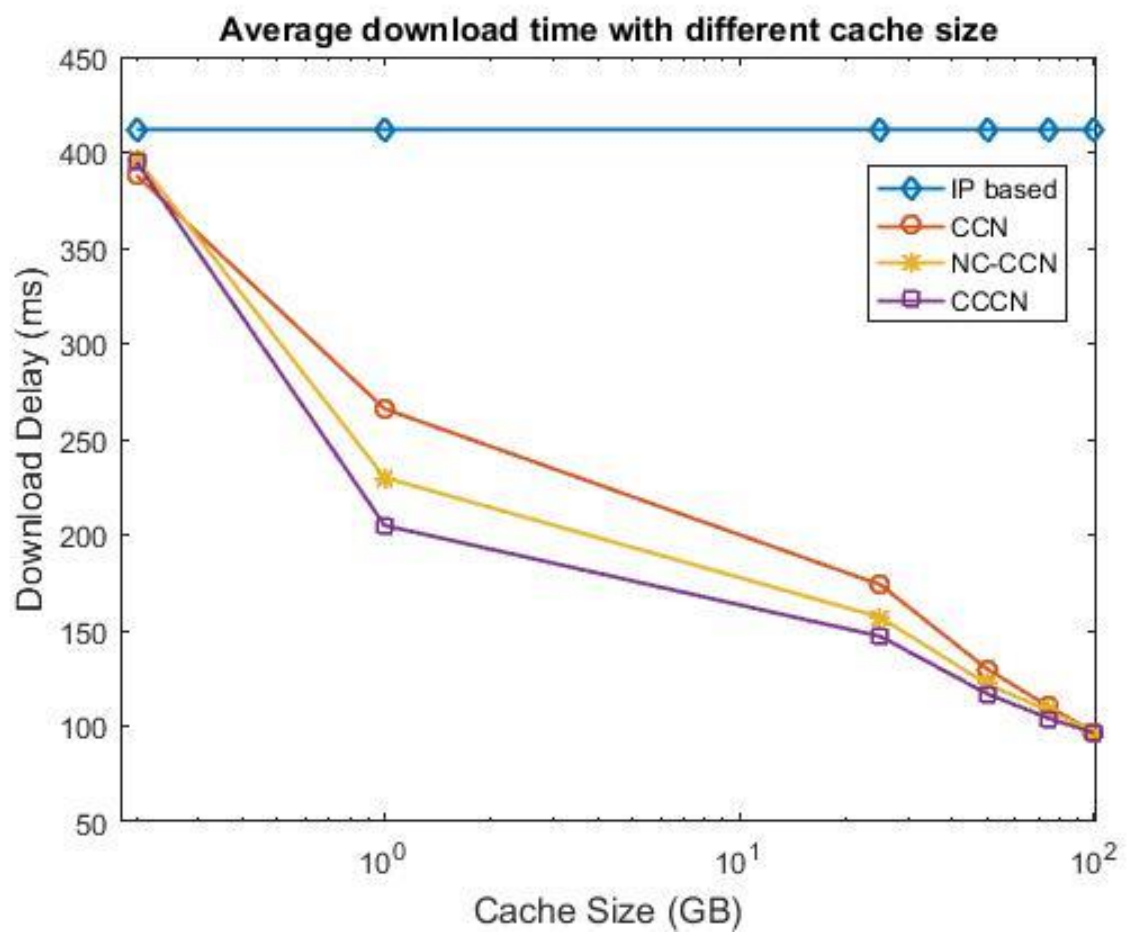

Fig. 6. Average download delay observed at gateway-cache router for different cache sizes.

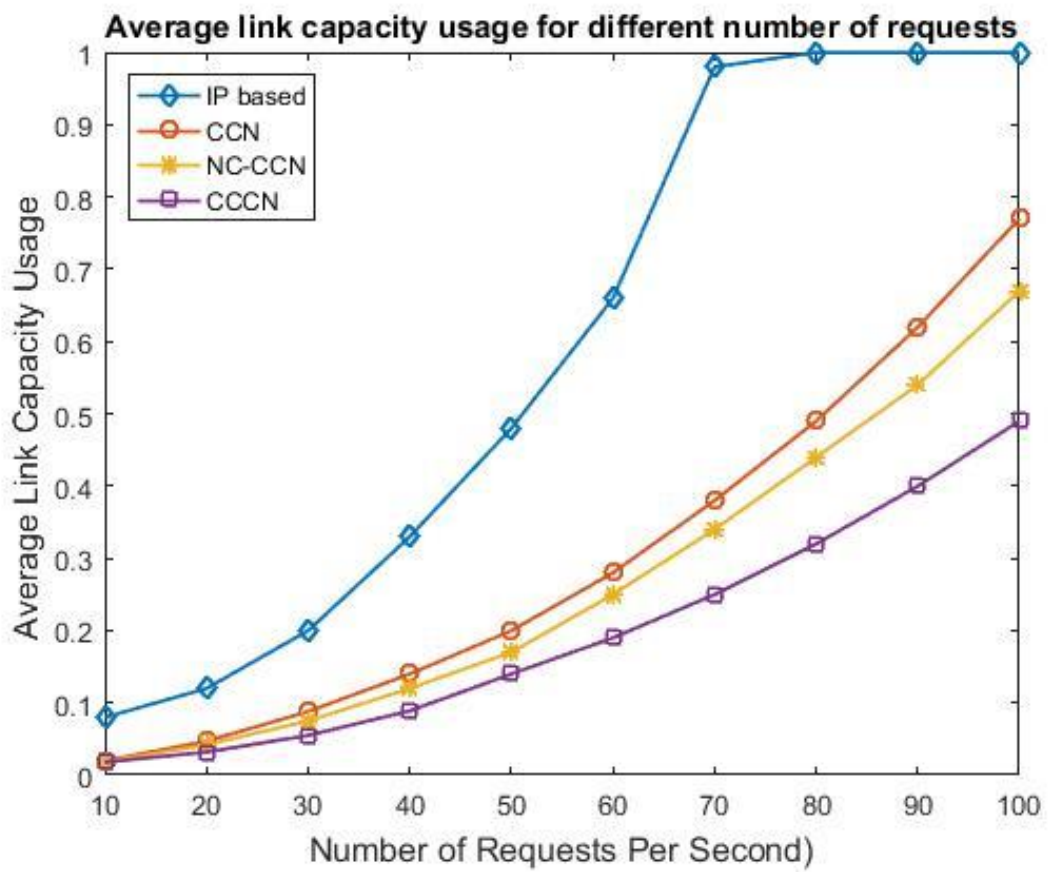

Fig. 7. Average link-capacity usage for different request loads.

Fig. 7 shows the comparison of the average link-capacity usage observed when each of the 15 gateway-cache routers, with a fixed cache size of 1 GB, receives the requests for the contents assumed to be Poisson distributed with a rate $\lambda_j^d = (10{\sim}100)\, req/s$.

**Definition:** In the steady state, the average link load is defined as the ratio of the total network traffic to the total number of links involved in the transmission. Thus, the link-capacity usage is the ratio of the average link load to the total link capacity.

The results are taken for the request arrival rate $\lambda_j^d = (10{\sim}100)\, req/s$. Fig. 7 clearly shows that both the CCN and CCCN perform better than the IP-based network. However, NC-CCN and CCCN performs better than the CCN, and the difference in the performances increases with the increase in the number of requests. This observation is quite evident because both, NC-CCN and CCCN, linearly combines the multiple data packets for high network traffic, thus reducing the usage of the link capacity.

In NC-CCN the interest packet carries the rank of coding vector. When a cache router receives an interest packet, it serves the request only if the rank value in interest packet is smaller than the number of saved coded blocks. This method potentially makes some false decisions; causing the wastage of network transmission and caching resources. In contrast, in CCCN the interest packet carries required coding information in $I_{Type}$ and $I_{info}$, and there is no chance of false decision; hence no wastage of network transmission and caching resources. Therefore, it clear from results, as shown in Fig. 7, NC-CCN link capacity usage is higher than CCCN.

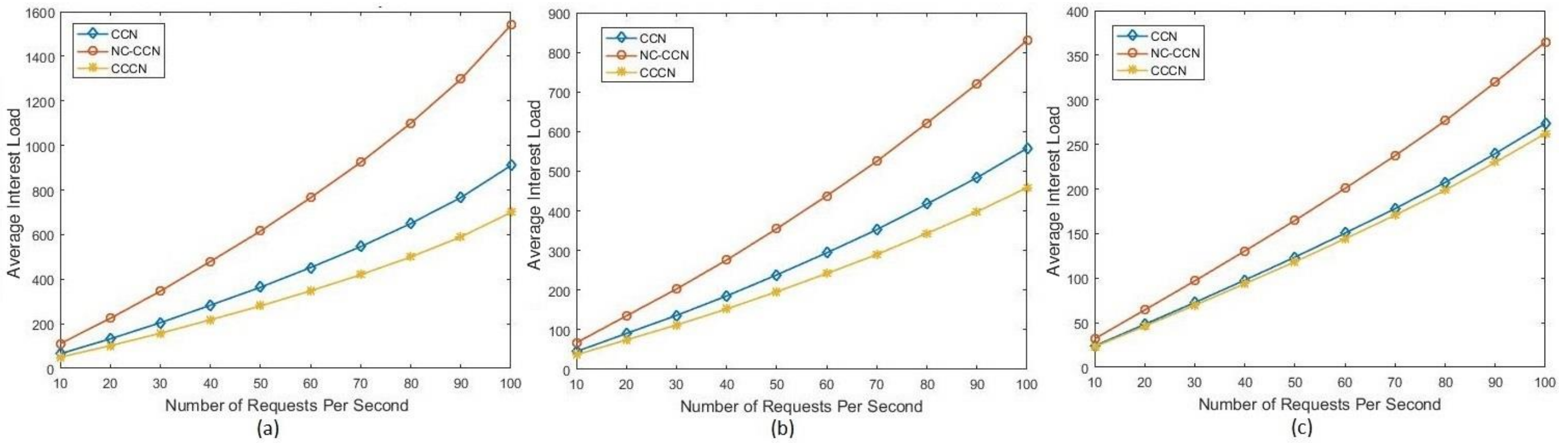

Fig. 8. Average number of interests for different request loads: (a) 1GB cache size (b) 10GB cache size (c) 100GB cache size.

In NC-CCN the interest packet carries the rank of coding vector. When a cache router receive an interest packet, it serves the request only if the rank value in interest packet is smaller than number of saved coded blocks. This method potentially can make false decisions; causing the wastage of network transmission and caching resources. In contrast, in CCCN the interest packet carries required coding information in $I_{Type}$ and $I_{info}$, and there is no chance of false decision; hence no wastage of network transmission and caching resources. Therefore, it clear from results, as shown in Fig. 7, NC-CCN link capacity usage is higher than CCCN.

Fig. 8 shows the comparison of the average number of interests per link observed when each of the 15 gateway-cache routers, with a cache size of 1GB, 10, and 100GB, receives the requests for the contents assumed to be Poisson distributed with a rate $\lambda_j^d = (10{\sim}100)\ req/s$.

**Definition:** In the steady state, the average interest load is defined as the ratio of the total number of interests forwarded in core network to the total number of requested content received at the gateway routers.

Fig. 8-a~c shows that both the CCN and CCCN perform better than the NC-CCN. The performance gap with cache size 1GB is significantly higher than the performance gap with cache size 10GB and cache size 100GB, as shown in Fig. 8-a, Fig. 8-b and Fig 8-c, respectively. The difference in the performances increases with the increase of request rate; however, for larger cache sizes, the difference in performance gap reduces. This observation is quite evident, because in NC-CCN the PIT is ineffective due to the content level naming. One of the key advantages of PIT is the aggregation of interest packets, due to the ineffectiveness of PIT in NC-CCN, the number of interest packets increases in proportion to the number of coded segments and the number of requests. Besides the performance degradation, the ineffectiveness of PIT also makes NC-CCN scheme vulnerable to the denial of service (DoS) attacks.

## VI. Conclusion

The ICN is a network-level solution, fundamentally considered a multisource, multicast content-delivery solution. Because of the benefits of network coding for multicasting scenarios and proven benefits for distributed storage networks, the network coding is apt for the ICN architecture. In this study, first, we showed that for the ICN architecture, a linear coding scheme exists, which is solvable in polynomial time. We adopted the model and approach discussed in another study [11] and presented an algebraic framework. Based on the theoretical algebraic framework, we proposed a practical implementation of the network-coding scheme for the ICN, particularly for the CCN architecture, termed the CCCN. We compared the proposed CCCN scheme with the NC-CCN, CCN, and IP-based network in terms of three performance parameters: the average download time, the average link-capacity usage, and the average interest load. The results proved that the network coding significantly improves the performance of the CCN; however, CCCN outperforms NC-CCN [30] in all cases.

## References

[1] G. Pallis and A. Vakali, "Insight and Perspectives for Content Delivery Networks," *Communications of the ACM*, vol. 49, no. 1, Jan 2006.

[2] C. Ge, Z. Sun, and N. Wang, "A Survey of Power-Saving Techniques on Data Centers and Content Delivery Networks," *IEEE Communications Surveys and Tutorials*, vol. 15, no. 3, 3rd Quarter 2013.

[3] S. Podlipnig and L. Boszormenyi, "A Survey of Web Cache Replacement Strategies,"*ACM Computing Surveys*, vol. 35, no. 4, Dec. 2003.

[4] J. Wang, "A Survey of Web Caching Schemes for the Internet," *ACM SIGCOMM Computer Communication Review*, vol. 29, no.5, Oct. 1999.

[5] E. K. Lua, J. Crowcroft, and M. Pias, "A survey and Comparison of Peer-To-Peer Overlay Network Schemes," *IEEE Communications Surveys*, vol. 7, no. 2, 2nd Quarter 2005.

[6] A. Malatras, "State-of-the-art survey on P2P overlay networks in pervasive computing environments," *Journal of Network and Computer Applications*, vol. 55, Sept. 2015.

[7] V. Jacobson, D. K. Smetters, J. D. Thornton, M. F. Plass, N, H. Briggs, and R. L. Braynard, "Networking Named Content," in *Proc. ACM CoNEXT '09*, Rome, Italy, Dec. 2009.

[8] B. Ahlgren, C. Dannewitz, C. Imbrenda, D. Kutscher, and B. Ohlman, "A Survey of Information-Centric Networking," *IEEE Communications Magazine*, vol. 50, no. 7, July 2012.

[9] G. Zhang, Y. Li, and T. Lin, "Caching in information centric networking: A survey," *Computer Networks*, vol. 57, no. 16, Nov. 2013.

[10] M. Bilal and S. G. Kang, "Time Aware Least Recent Used (TLRU) Cache Management Policy in ICN," in *Proc. IEEE ICACT'14* Pyeongchang, Korea (South), Feb. 2014.
[11] R. Koetter and M. Medrd, "An algebraic approach to network coding," *IEEE/ACM Transactions on Networking*, vol.11, no. 5, Oct. 2003.
[12] C. Fragouli and E. Soljanin, "Network coding fundamentals," Foundations and Trends® in Networking, vol. 2, no. 1, Jan. 2007.
[13] X.Yan, J. Yang and Z. Zhang, "An Outer Bound for Multisource Multisink Network Coding With Minimum Cost Consideration," *IEEE Transactions on Information Theory*, vol. 52, no. 6, Jun. 2006.
[14] Al. Cohen, As. Cohen, M. Medrd, O. Gurewitz, "Secure Multi-Source Multicast," arXiv:1702.03012v1 [cs.IT], Feb. 2017.
[15] M. Kwon and H. Park, "Network coding-based distributed network formation game for multi-source multicast networks," in *Proc. IEEE ICC'17*, Paris, France, May 2017.
[16] J. Choi, J. Han, E. Cho, T.Kwon, and Y. Choi, "A Survey on Content-Oriented Networking for Efficient Content Delivery," IEEE Communications Magazine, vol. 49, no. 3, Mar. 2011.
[17] A. G. Dimakis, P. B. Godfrey, Y. Wu, M. Wainwright and K. Ramchandran, "Network Coding for Distributed Storage Systems," *IEEE Transactions on Information Theory*, vol. 56, no. 9, Jun. 2010.
[18] M. Yang and Y. Yang, "Applying Network Coding to Peer-to-Peer File Sharing," *IEEE Transactions on Computers,* vol. 63, no. 8, Aug. 2014.
[19] B. Li and D. Niu, "Random Network Coding in Peer-to-Peer Networks: From Theory to Practice," *Proceedings of the IEEE*, vol. 99, no. 3, Mar. 2011.
[20] H. Ayatollahi, M. Khansari, and H. R. Rabiee, "A push-pull network coding protocol for live peer-to-peer streaming," *Computer Networks,* vol. 130, no. C, Jan. 2018.
[21] C. Gkantsidis and P. R. Rodriguez, "Network Coding for Large Scale Content Distribution," in *Proc. IEEE Infocom '05,* Florida, USA, Mar. 2005.
[22] M. Montpetit, C. Westphal and D. Trossen, "Network Coding Meets Information-Centric Networking: An Architectural Case for Information Dispersion Through Native Network Coding," *in Proc. NoM'12*, South Carolina, USA, Jun. 2012.
[23] J. Llorca, A. M. Tulino, K. Guan and D. C. Kilper, "Network-Coded Caching-Aided Multicast for Efficient Content Delivery," *in Proc. IEEE ICC'13,* Budapest, Hungary, Jun. 2013.
[24] Q. Wu and Z. Li, G. Xie, "CodingCache: Multipath-aware CCN Cache with Network Coding," in Proc. ICN'13, Hong-Kong, China, Aug. 2013.
[25] G. Parisis and D. Torssen, "Filling the Gaps of Unused Capacity through a Fountain Coded Dissemination of Information," *Mobile Computing and Communications Review,* vol.18, no. 1, Jan. 2014.
[26] D. Torssen and G. Parisis, "Designing and Realizing an Information-Centric Internetm," *IEEE Communications Magazine,* vol.50,no. 7, Jul. 2012.
[27] T. Koponen et al., "A Data-Oriented Network Architecture," *ACM SIGCOMM*, 2007.
[28] J. W. Byers, M. Luby, M. Mitzenmacher, and A. Rege, "A digital fountain approach to reliable distribution of bulk data," *In Proc. SIGCOMM 98*, pages 56–67, 1998.
[29] G. Parisis et. al., "Efficient content delivery through fountain coding in opportunistic information-centric networks," *Computer Communications,* vol. 100, Mar. 2017.
[30] G. Zhang and Z. Xu, "Combing CCN with network coding: An architectural perspective," *Computer Networks*, vol. 94, Jan. 2016.
[31] K. Matsuzono, H. Asaeda, and T. Turletti, "Low Latency Low Loss Streaming using In-Network Coding and Caching," in *Proc. IEEE Infocom '17*, Atlanta, USA, May 2017.
[32] R. Ahlswede, N. Cai, S.-Y, Li, and R. Yeung, "Network information flow," *IEEE Trans. Information Theory*, vol. 46, no. 4, July 2000.
[33] R. Boussaha, Y. Challal, M. Bessedik, and A. Bouabdallah, "Towards Authenticated Network Coding for Named Data Networking," in Proc. of 24th International Conference on Software, Telecommunications and Computer Networks (SoftCom '17), Split, Croatia, Sep. 2017.
[34] T. Ho, M. Médard, J. Shi, M. Effros, and D. R. Karger, "On randomized network coding," in *Proc. 41st Annu. Allerton Conf. Communication, Control, and Computing,* Monticello, IL, Oct. 2003.
[35] D. Traskov, N. Ratnakar, D. S. Lun, R. Koetter, and M. Medard, "Network Coding for Multiple Unicasts: An Approach based on Linear Optimization,"in *Proc. IEEE International Symposium on Information Theory '06,* WA, USA, Jul. 2006.
[36] Tracey Ho, M. Medard, J. Shi, M. Effros and D. R. Karger, "A Random Linear Network Coding Approach to Multicast," *IEEE Transactions on Information Theory*, vol. 52, no. 10, Jun. 2006.
[37] C. Chiasserini, E.Viterbo and C. Casetti, "Decoding Probability in Random Linear Network Coding with Packet Losses," *IEEE Communications Letters*, vol. 17, no. 11, Nov. 2013.
[38] L. Muscariello, "Content-Centric Networking Packet Header Format draft-ccn-packet-header-00," IETF, Internet-Draft, raft-ccn-packet-header-00, Jul. 2015.
[39] M. Bilal and S. Kang, "A Cache Management Scheme for Efficient Content Eviction and Replication in Cache Networks," IEEE Access, vol. 5, no. 1, pp. 1962–1701, Dec. 2017.

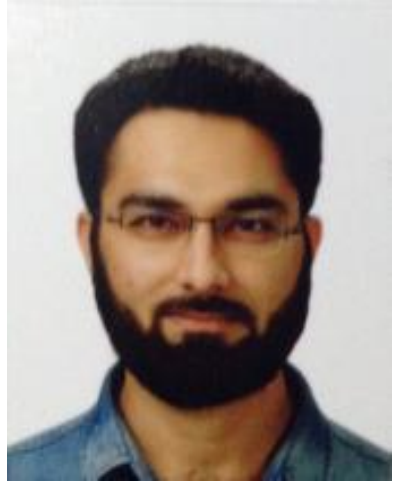

**Muhammad Bilal** is an assistant professor of computer science in the Department of Computer and Electronic Systems Engineering at Hankuk University of Foreign Studies, Yongin, Rep. of Korea. He received his Ph.D. degree in Information and Communication Network Engineering from Korea University of Science and Technology, school of Electronics and Telecommunications Research Institute (ETRI), MS in computer engineering from Chosun University, Gwangju, Rep. of Korea, and BS degree in computer systems engineering from University of Engineering and Technology, Peshawar, Pakistan. Prior to joining Hankuk University of Foreign Studies, he was a postdoctoral research fellow at Smart Quantum Communication Center, Korea University. He has served as a reviewer of various international Journals including IEEE Systems Journal, IEEE Access, IEEE Communications Letters, IEEE Transactions on Network and Service Management, Journal of Network and Computer Applications, Personal and Ubiquitous Computing and International Journal of Communication Systems. He has also served as a program committee member on many international conferences. His primary research interests are Design and Analysis of Network Protocols, Network Architecture, Network Security, IoT, Named Data Networking, Cryptology and Future Internet.

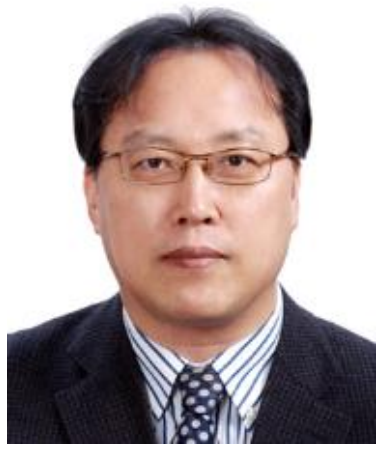

**Shin-Gak Kang** received his BS and MS degree in electronics engineering from Chungnam National University, Rep. of Korea, in 1984 and 1987, respectively and his Ph.D. degree in information communication engineering from Chungnam National University, Rep of Korea in 1998. Since 1984, he is working with Electronics and Telecommunications Research Institute, Daejeon, Rep. of Korea, where he is a Director of Open Source Center of ETRI. From 2008 he is a professor at the Department of Information and Communication Network Technology, University of Science and Technology, Korea. He is actively participating in various international standardization activities as a Vice-chairman of ITU-T SG11, Chairman of WP2/11, and Convenor of ISO/IEC JTC 1/SC 6/WG 7, etc. His research interests include Future Network, IMT-2020 network, Information Centric Networking, multimedia communications and Applications, and various ICT converged services.